\documentstyle[preprint,tighten,eqsecnum,floats,aps]{revtex}

\def\be{\begin{equation}}
\def\ee{\end{equation}}
\def\ba{\begin{eqnarray}}
\def\ea{\end{eqnarray}}
\def\g{\gamma}
\def\Pl{\ell_P}

\def\jd{j^{(d)}}
\def\ju{j^{(u)}}
\def\jdu{j^{(d+u)}}

\def\H{{\cal H}}
\def\Hg{{\cal H}_\gamma}

\preprint{\vbox{\baselineskip=12pt
\rightline{CGPG-98/12-5}}}

\begin{document}
\draft
\title{Quantum Mechanics of Geometry}
\author {A.\ Ashtekar
\thanks{It is a pleasure to dedicate this article to Professor Jayant
Narlikar on the occasion of his 60th birthday.}}
 
\address{Center for Gravitational Physics and Geometry \\
Department of Physics, The Pennsylvania State University \\
University Park, PA 16802, USA}

\maketitle 


\begin{abstract}

Over the past six years, a detailed framework has been constructed to
unravel the quantum nature of the Riemannian geometry of physical
space.  A review of these developments is presented at a level which
should be accessible to graduate students in physics. As an
illustrative application, I indicate how some of the detailed features
of the micro-structure of geometry can be tested using black hole
thermodynamics. Current and future directions of research in this
area are discussed.

\end{abstract}

\section{Introduction}
\label{s1}

During his G\"ottingen inaugural address in 1854, Riemann [1]
suggested that geometry of space may be more than just a fiducial,
mathematical entity serving as a passive stage for physical phenomena,
and may in fact have direct physical meaning in its own right.
General relativity provided a brilliant confirmation of this vision:
curvature of space now encodes the physical gravitational field. This
shift is profound. To bring out the contrast, let me recall the
situation in Newtonian physics. There, space forms an inert arena on
which the dynamics of physical systems --such as the solar system--
unfolds. It is like a stage, an unchanging backdrop for all of
physics. In general relativity, by contrast, the situation is very
different.  Einstein's equations tell us that matter curves
space. Geometry is no longer immune to change. It reacts to matter. It
is dynamical.  It has ``physical degrees of freedom'' in its own
right. In general relativity, the stage disappears and joins the
troupe of actors! Geometry is a physical entity, very much like
matter.

Now, the physics of this century has shown us that matter has
constituents and the 3-dimensional objects we perceive as solids are
in fact made of atoms. The continuum description of matter is an
approximation which succeeds brilliantly in the macroscopic regime but
fails hopelessly at the atomic scale. It is therefore natural to ask:
Is the same true of geometry? If so, what is the analog of the `atomic
scale?' We know that a quantum theory of geometry should contain three
fundamental constants of Nature, $c, G, \hbar$, the speed of light,
Newton's gravitational constant and Planck's constant.  Now, as Planck
pointed out in his celebrated paper that marks the beginning of
quantum mechanics, there is a unique combination, $\ell_P =\sqrt
{\hbar G/c^3}$, of these constants which has dimension of
length. ($\ell_P \approx 10^{-33}$cm.) It is now called the Planck
length. Experience has taught us that the presence of a distinguished
scale in a physical theory often marks a potential transition; physics
below the scale can be very different from that above the scale. Now,
all of our well-tested physics occurs at length scales much bigger
than than $\ell_P$. In this regime, the continuum picture works
well. A key question then is: Will it break down at the Planck length?
Does geometry have constituents at this scale? If so, what are its
atoms?  Its elementary excitations?  Is the space-time continuum only
a `coarse-grained' approximation? Is geometry quantized?  If so, what
is the nature of its quanta?

To probe such issues, it is natural to look for hints in the
procedures that have been successful in describing matter. Let us
begin by asking what we mean by quantization of physical quantities.
Take a simple example --the hydrogen atom. In this case, the answer is
clear: while the basic observables --energy and angular momentum--
take on a continuous range of values classically, in quantum mechanics
their eigenvalues are discrete; they are quantized. So, we can ask if
the same is true of geometry. Classical geometrical quantities such as
lengths, areas and volumes can take on continuous values on the phase
space of general relativity. Are the eigenvalues of corresponding
quantum operators discrete? If so, we would say that geometry is
quantized and the precise eigenvalues and eigenvectors of geometric
operators would reveal its detailed microscopic properties.

Thus, it is rather easy to pose the basic questions in a precise
fashion. Indeed, they could have been formulated soon after the advent
of quantum mechanics. Answering them, on the other hand, has proved to
be surprisingly difficult. The main reason, I believe, is the
inadequacy of standard techniques. More precisely, to examine the
microscopic structure of geometry, we must treat Einstein gravity
quantum mechanically, i.e., construct at least the basics of a quantum
theory of the gravitational field. Now, in the traditional approaches
to quantum field theory, one {\it begins} with a continuum, background
geometry. To probe the nature of quantum geometry, on the other hand,
we should {\it not} begin by assuming the validity of this picture. We
must let quantum gravity decide whether this picture is adequate; the
theory itself should lead us to the correct microscopic model of
geometry.

With this general philosophy, in this article I will summarize the
picture of quantum geometry that has emerged from a specific approach
to quantum gravity. This approach is non-perturbative. In perturbative
approaches, one generally begins by assuming that space-time geometry
is flat and incorporates gravity --and hence curvature-- step by step
by adding up small corrections. Discreteness is then hard to unravel%
\footnote{The situation can be illustrated by a harmonic oscillator:
While the exact energy levels of the oscillator are discrete, it would
be very difficult to ``see'' this discreteness if one began with a
free particle whose energy levels are continuous and then tried to
incorporate the effects of the oscillator potential step by step via
perturbation theory.}.
In the non-perturbative approach, by contrast, there is no background
metric at all. All we have is a bare manifold to start with. All
fields --matter as well as gravity/geometry-- are treated as dynamical
from the beginning. Consequently, the description can not refer to a
background metric. Technically this means that the full diffeomorphism
group of the manifold is respected; the theory is generally covariant.

As we will see, this fact leads one to Hilbert spaces of quantum
states which are quite different from the familiar Fock spaces of
particle physics. Now gravitons --the three dimensional wavy
undulations on a flat metric-- do not represent fundamental
excitations. Rather, the fundamental excitations are {\it one}
dimensional. Microscopically, geometry is rather like a
polymer. Recall that, although polymers are intrinsically one
dimensional, when densely packed in suitable configurations they can
exhibit properties of a three dimensional system. Similarly, the
familiar continuum picture of geometry arises as an approximation: one
can regard the fundamental excitations as `quantum threads' with which
one can `weave' continuum geometries. That is, the continuum picture
arises upon coarse-graining of the semi-classical `weave states'.
Gravitons are no longer the fundamental mediators of the gravitational
interaction. They now arise only as approximate notions.  They
represent perturbations of weave states and mediate the gravitational
force only in the semi-classical approximation. Because the
non-perturbative states are polymer-like, geometrical observables turn
out to have discrete spectra. They provide a rather detailed picture
of quantum geometry from which physical predictions can be made.

The article is divided into two parts. In the first, I will indicate how
one can reformulate general relativity so that it resembles gauge
theories. This formulation provides the starting point for the quantum
theory. In particular, the one-dimensional excitations of geometry
arise as the analogs of `Wilson loops' which are themselves analogs of
the line integrals $\exp i\oint A.d\ell$ of electro-magnetism. In the
second part, I will indicate how this description leads us to a
quantum theory of geometry. I will focus on area operators and show
how the detailed information about the eigenvalues of these operators
has interesting physical consequences, e.g., to the process of Hawking
evaporation of black holes.

I should emphasize that this is {\it not} a technical review. Rather,
it is written in the same spirit that drives Jayant's educational
initiatives. I thought this would be a fitting way to honor Jayant
since these efforts have occupied so much of his time and energy in
recent years.  Thus my aim is present to beginning researchers an
overall, semi-quantitative picture of the main ideas. Therefore, the
article is written at the level of colloquia in physics departments in
the United States.  I will also make some historic detours of general
interest.  At the end, however, I will list references where the
details of the central results can be found.

\section{From metrics to connections}

\subsection{Gravity versus other fundamental forces}

General relativity is normally regarded as a dynamical theory of
metrics ---tensor fields that define distances and hence geometry.  It
is this fact that enabled Einstein to code the gravitational field in
the Riemannian curvature of the metric. Let me amplify with an
analogy. Just as position serves as the configuration variable in
particle dynamics, the three dimensional metric of space can be taken
to be the configuration variable of general relativity. Given the
initial position and velocity of a particle, Newton's laws provide us
with its trajectory in the position space. Similarly, given a three
dimensional metric and its time derivative at an initial instant,
Einstein's equations provide us with a four dimensional space-time
which can be regarded as a trajectory in the space of 3-metrics
\footnote{Actually, only six of the ten Einstein's equations provide 
the evolution equations. The other four do not involve
time-derivatives at all and are thus constraints on the initial values of
the metric and its time derivative. However, if the constraint equations
are satisfied initially, they continue to be satisfied at all times.}.

However, this emphasis on the metric sets general relativity apart
from all other fundamental forces of Nature. Indeed, in the theory of
electro-weak and strong interactions, the basic dynamical variable is
a (matrix-valued) vector potential, or a connection. Like general
relativity, these theories are also geometrical. The connection
enables one to parallel-transport objects along curves. In
electrodynamics, the object is a charged particle such as an electron;
in chromodynamics, it is a particle with internal color, such as a
quark. Generally, if we move the object around a closed loop, we find
that its state does not return to the initial value; it is rotated by
an unitary matrix.  In this case, the connection is said to have
curvature and the unitary matrix is a measure of the curvature in a
region enclosed by the loop. In the case of electrodynamics, the
connection is determined by the vector potential and the curvature by
the electro-magnetic field strength.

Since the metric also gives rise to curvature, it is natural to ask if
there is a relation between metrics and connections. The answer is in
the affirmative. Every metric defines a connection ---called the
Levi-Civita connection of the metric. The object that the connection
enables one to parallel transport is a vector. (It is this connection
that determines the geodesics, i.e. the trajectories of particles in
absence of non-gravitational forces.) It is therefore natural to ask
if one can not use this connection as the basic variable in general
relativity. If so, general relativity would be cast in a language that
is rather similar to gauge theories and the description of the
(general relativistic) gravitational interaction would be very similar
to that of the other fundamental interactions of Nature. It turns out
that the answer is in the affirmative. Furthermore, both Einstein and
Schr\"odinger gave such a reformulation of general relativity.  Why is
this fact then not generally known? Indeed, I know of no textbook on
general relativity which even mentions it. One reason is that in their
reformulation the basic equations are somewhat complicated ---but not
much more complicated, I think, than the standard ones in terms of the
metric. A more important reason is that we tend to think of distances,
light cones and causality as fundamental. These are directly
determined by the metric and in a connection formulation, the metric
is a `derived' rather than a fundamental concept. But in the last few
years, I have come to the conclusion that the real reason why the
connection formulation of Einstein and Schr\"odinger has remained so
obscure may lie in an interesting historical episode. I will return to
this point at the end of this section.

\subsection{Metrics versus connections}

Modern day researchers re-discovered connection theories of gravity
after the invention and successes of gauge theories for other
interactions. Generally, however, these formulations lead one to
theories which are quite distinct from general relativity and the
stringent experimental tests of general relativity often suffice to
rule them out. There is, however, a reformulation of general
relativity itself in which the basic equations are simpler than the
standard ones: while Einstein's equations are non-polynomial in terms
of the metric and its conjugate momentum, they turn out to be low
order polynomials in terms of the new connection and its conjugate
momentum.  Furthermore, just as the simplest particle trajectories in
space-time are given by geodesics, the `trajectory' determined by the
time evolution of this connection according to Einstein's equation
turns out to be a geodesic in the configuration space of connections.

In this formulation, the phase space of general relativity is
identical to that of the Yang-Mills theory which governs weak
interactions. Recall first that in electrodynamics, the (magnetic)
vector potential constitutes the configuration variable and the
electric field serves as the conjugate momentum. In weak interactions
and general relativity, the configuration variable is a matrix-valued
vector potential; it can be written as $\vec{A}_i\tau_i$ where
$\vec{A}_i$ is a triplet of vector fields and $\tau_i$ are the Pauli
matrices. The conjugate momenta are represented by $\vec{E}_i\tau_i$
where $\vec{E}_i$ is a triplet of vector fields%
\footnote{As usual, summation over the repeated index $i$ is assumed. 
Also, technically each $\vec{A}_i$ is a 1-form rather than a vector
field.  Similarly, each $\vec{E}_i$ is a vector density of weight one,
i.e., natural dual of a 2-form.}.
Given a pair $(\vec{A}_i, \vec{E}_i)$ (satisfying appropriate
conditions as noted in footnote 2), the field equations of the two
theories determine the complete time-evolution, i.e., a dynamical
trajectory. 

The field equations --and the Hamiltonians governing them-- of the two
theories are of course very different. In the case of weak
interactions, we have a background space-time and we can use its
metric to construct the Hamiltonian. In general relativity, we do not
have a background metric. On the one hand this makes life very
difficult since we do not have a fixed notion of distances or causal
structures; these notions are to arise from the solution of the
equations we are trying to write down! On the other hand, there is
also tremendous simplification: Because there is no background metric,
there are very few mathematically meaningful, gauge invariant
expressions of the Hamiltonian that one can write down. (As we will
see, this theme repeats itself in the quantum theory.)  It is a
pleasant surprise that the simplest non-trivial expression one can
construct from the connection and its conjugate momentum is in fact
the correct one, i.e., is the Hamiltonian of general relativity! The
expression is at most quadratic in $\vec{A}_i$ and at most quadratic
in $\vec{E}_i$. The similarity with gauge theories opens up new
avenues for quantizing general relativity and the simplicity of the
field equations makes the task considerably easier.

What is the physical meaning of these new basic variables of general
relativity? As mentioned before, connections tell us how to parallel
transport various physical entities around curves. The Levi-Civita
connection tells us how to parallel transport vectors. The new
connection, $\vec{A}_i$, on the other hand, determines the parallel
transport of {\it left handed spin}-$\frac{1}{2}$ {\it particles}
(such as the fermions in the standard model of particle physics)
---the so called {\it chiral fermions}. These fermions are
mathematically represented by spinors which, as we know from
elementary quantum mechanics, can be roughly thought of as `square
roots of vectors'. Not surprisingly, therefore, the new connection is
not completely determined by the metric alone. It requires additional
information which roughly is a square-root of the metric, or a tetrad.
The conjugate momenta $\vec{E}_i$ represent restrictions of these
tetrads to space. They can be interpreted as spatial triads, i.e., as
`square-roots' of the metric of the 3-dimensional space.  Thus,
information about the Riemannian geometry of space is coded directly
in these momenta. The (space and) time-derivatives of the triads are
coded in the connection.

To summarize, there {\it is} a formulation of general relativity which
brings it closer to theories of other fundamental
interactions. Furthermore, in this formulation, the field equations
simplify greatly. Thus, it provides a natural point of departure for 
constructing a quantum theory of gravity and for probing the nature of
quantum geometry non-perturbatively.

\subsection{Historical detour}

To conclude this section, let me return to the piece of history
involving Einstein and Schr\"odinger that I mentioned earlier. In the
forties, both men were working on unified field theories. They were
intellectually very close. Indeed, Einstein wrote to Schr\"odinger
saying that he was perhaps the only one who was not `wearing blinkers'
in regard to fundamental questions in science and Schr\"odinger
credited Einstein for inspiration behind his own work that led to the
Schr\"odinger equation. Einstein was in Princeton and Schr\"odinger in
Dublin. But During the years 1946-47, they frequently exchanged ideas
on unified field theory and, in particular, on the issue of whether
connections should be regarded as fundamental or metrics.  In fact the
dates on their letters often show that the correspondence was going
back and forth with astonishing speed. It reveals how quickly they
understood the technical material the other hand sent, how they
hesitated, how they teased each other. Here are a few quotes:
\medskip

\noindent
{\sl The whole thing is going through my head like a millwheel: To
take $\Gamma$} [the connection] {\sl alone as the primitive variable
or the $g$'s} [metrics] {\sl and $\Gamma$'s ? ...}\hfil\break
\indent\indent ---Schr\"odinger, May 1st, 1946.
\medskip

\noindent
{\sl How well I understand your hesitating attitude! I must confess to
you that inwardly I am not so certain ... We have squandered a lot of
time on this thing, and the results look like a gift from devil's
grandmother.}\hfil\break
\indent\indent ---Einstein, May 20th, 1946
\medskip

\noindent
Einstein was expressing doubts about using the Levi-Civita connection
alone as the starting point which he had advocated at one time.
Schr\"odinger wrote back that he laughed very hard at the phrase
`devil's grandmother'. In another letter, Einstein called
Schr\"odinger `a clever rascal'. Schr\"odinger was delighted and
took it to be a high honor.  This continued all through 1946. Then, in
the beginning of 1947, Schr\"odinger thought he had made a
breakthrough. He wrote to Einstein:
\medskip

\noindent
{\sl Today, I can report on a real advance. May be you will grumble
frightfully for you have explained recently why you don't approve of my
method. But very soon, you will agree with me...}\hfil\break
\indent\indent ---Schr\"odinger, January 26th, 1947
\medskip

\noindent
Schr\"odinger sincerely believed that his breakthrough was
revolutionary
\footnote{The `breakthrough' was to drop the requirement that the 
(Levi-Civita) connection be symmetric, i.e., to allow for torsion.}.
Privately, he spoke of a second Nobel prize. The very next day after
he wrote to Einstein, he gave a seminar in the Dublin Institute of
Advanced Studies. Both the Taoiseach (the Irish prime minister) and
newspaper reporters were invited. The day after, the following
headlines appeared:
\medskip

\noindent
{\sl Twenty persons heard and saw history being made in the world of
physics.  ... The Taoiseach was in the group of professors and
students. ..}[To a question from the reporter] {\sl Professor
Schr\"odinger replied ``This is the generalization. Now the Einstein
theory becomes simply a special case ...''}\hfil\break \indent\indent
---Irish Press, January 28th, 1947
\medskip

\noindent
Not surprisingly, the headlines were picked up by New York Times which
obtained photocopies of Schr\"odinger's paper and sent them to
prominent physicists --including of course Einstein-- for comments. As
Walter Moore, Schr\"odinger's biographer puts it, Einstein could
hardly believe that such grandiose claims had been made based on a
what was at best a small advance in an area of work that they both had
been pursuing for some time along parallel lines. He prepared a
carefully worded response to the request from New York
Times:\hfil\break

\noindent
{\sl It seems undesirable to me to present such preliminary attempts
to the public. ... Such communiqu\'es given in sensational terms give
the lay public misleading ideas about the character of research. The
reader gets the impression that every five minutes there is a
revolution in Science, somewhat like a coup d'\'etat in some of the
smaller unstable republics. ...}
\medskip

Einstein's comments were also carried by the international press.  On
seeing them, Schr\"odinger wrote a letter of apology to Einstein
citing his desire to improve the financial conditions of physicists in
the Dublin Institute as a reason for the exaggerated account. It seems
likely that this `explanation' only worsened the situation. Einstein
never replied. He also stopped scientific communication with
Schr\"odinger for three years.

The episode must have been shocking to those few who were exploring
general relativity and unified field theories at the time. Could it be
that this episode effectively buried the desire to follow up on
connection formulations of general relativity until an entirely new
generation of physicists who were blissfully unaware of this episode
came on the scene?

\section{Quantum Geometry}

\subsection{General Setting}

Now that we have a connection formulation of general relativity, let
us consider the problem of quantization. Recall first that in the
quantum description of a particle, states are represented by suitable
wave functions $\Psi(\vec{x})$ on the classical configuration space of
the particle. Similarly, quantum states of the gravitational field are
represented by appropriate wave functions $\Psi(\vec{A}_i)$ of
connections. Just as the momentum operator in particle mechanics is
represented by $\hat{P}\cdot\Psi_{I} = -i\hbar\, (\partial\Psi/
\partial x_I)$ (with $I=1,2,3$), the triad operators are represented
by $\hat{\vec{E}_i}\cdot \Psi = -i\hbar G (\delta\Psi/ \delta{\vec
A}_i)$. The task is to express geometric quantities, such as lengths
of curves, areas of surfaces and volumes of regions, in terms of
triads using ordinary differential geometry and then promote these
expressions to well-defined operators on the Hilbert space of quantum
states. In principle, the task is rather similar to that in quantum
mechanics where we first express observables such as angular momentum
or Hamiltonian in terms of configuration and momentum variables
$\vec{x}$ and $\vec{p}$ and then promote them to quantum theory as
well-defined operators on the quantum Hilbert space.

In quantum mechanics, the task is relatively straightforward; the only
potential problem is the choice of factor ordering. In the present
case, by contrast, we are dealing with a {\it field theory}, i.e., a
system with an infinite number of degrees of freedom. Consequently, in
addition to factor ordering, we face the much more difficult problem
of regularization. Let me explain qualitatively how this arises. A
field operator, such as the triad mentioned above, excites infinitely
many degrees of freedom. Technically, its expectation values are
distributions rather than smooth fields. They don't take precise
values at a given point in space.  To obtain numbers, we have to
integrate the distribution against a test function, which extracts
from it a `bit' of information. As we change our test or smearing
field, we get more and more information.  (Take the familiar Dirac
$\delta$-distribution $\delta(x)$; it does not have a well-defined
value at $x= 0$. Yet, we can extract the full information contained in
$\delta(x)$ through the formula: $\int \delta(x) f(x) dx = f(0)$ for
all test functions $f(x)$.) Thus, in a precise sense, field operators
are distribution-valued. Now, as is well known, product of
distributions is not well-defined. If we attempt naively to give
meaning to it, we obtain infinities, i.e., a senseless
result. Unfortunately, all geometric operators involve rather
complicated (in fact non-polynomial) functions of the triads. So, the
naive expressions of the corresponding quantum operators are typically
meaningless. The key problem is to regularize these expressions, i.e.,
to extract well-defined operators from the formal expressions in a
coherent fashion.

This problem is not new; it arises in all physically interesting
quantum field theories. However, as I mentioned in the Introduction,
in other theories one has a background space-time metric and it is
invariably used in a critical way in the process of
regularization. For example, consider the electro-magnetic field.  We
know that the energy of the Hamiltonian of the theory is given by $H =
\int (\vec{E}\cdot\vec{E}+ \vec{B}\cdot\vec{B})\, d^3x$. Now, in the
quantum theory, $\hat{\vec{E}}$ and $\hat{\vec{B}}$ are both
operator-valued distributions and so their square is ill-defined. But
then, using the background flat metric, one Fourier decomposes these
distributions, identifies creation and annihilation operators and
extracts a well-defined Hamiltonian operator by normal ordering, i.e.,
by physically moving all annihilators to the right of creators. This
procedure removes the unwanted and unphysical infinite zero point
energy form the formal expression and the subtraction makes the
operator well-defined. In the present case, on the other hand, we are
trying to construct a quantum theory of geometry/gravity and do not
have a flat metric --or indeed, any metric-- in the background.
Therefore, many of the standard regularization techniques are no
longer available.

\subsection{Geometric operators}

Fortunately, between 1992 and 1995, a new functional calculus
was developed on the space of connections $\vec{A}_i$ ---i.e., on the
configuration space of the theory. This calculus is mathematically
rigorous and makes no reference at all to a background space-time
geometry; it is generally covariant.  It provides a variety of new
techniques which make the task of regularization feasible. First of
all, there is a well-defined integration theory on this space. To
actually evaluate integrals and define the Hilbert space of quantum
states, one needs a measure: given a measure on the space of
connections, we can consider the space of square-integrable functions
which can serve as the Hilbert space of quantum states. It turns out
that there is a preferred measure, singled out by the physical
requirement that the (gauge covariant versions of the) configuration
and momentum operators be self-adjoint. This measure is diffeomorphism
invariant and thus respects the underlying symmetries coming from
general covariance. Thus, there is a natural Hilbert space of states
to work with%
\footnote{ This is called the kinematical Hilbert space; it enables
one to formulate the quantum Einstein's (or supergravity)
equations. The final, physical Hilbert space will consist of states
which are solutions to these equations.}.
Let us denote it by $\H$. Differential calculus enables one to
introduce physically interesting operators on this Hilbert space and
regulate them in a generally covariant fashion. As in the classical
theory, the absence of a background metric is both a curse and a
blessing.  On the one hand, because we have very little structure to
work with, many of the standard techniques simply fail to carry over.
On the other hand, at least for geometric operators, the choice of
viable expressions is now severely limited which greatly simplifies the
task of regularization.

The general strategy is the following. The Hilbert space $\H$ is the
space of square-integrable functions $\Psi(\vec{A}_i)$ of connections
$\vec{A}_i$. A key simplification arises because it can be obtained as
the (projective) limit of Hilbert spaces associated with systems with
only a finite number of degrees of freedom. More precisely, given any
graph $\g$ (which one can intuitively think of as a `floating
lattice') in the physical space, using techniques which are very
similar to those employed in lattice gauge theory, one can construct a
Hilbert space $\Hg$ for a quantum mechanical system with $3N$ degrees
of freedom, where $N$ is the number of edges of the graph%
\footnote{The factor $3$ comes from the dimension of the gauge group 
${\rm SU}(2)$ which acts on Chiral spinors. The mathematical structure
of the gauge-rotations induced by this ${\rm SU}(2)$ is exactly the
same as that in the angular-momentum theory of spin-$\frac{1}{2}$
particles in elementary quantum mechanics.}.
Roughly, these Hilbert spaces know only about how the connection
parallel transports chiral fermions along the edges of the graph and
not elsewhere. That is, the graph is a mathematical device to extract
$3N$ `bits of information' from the full, infinite dimensional
information contained in the connection, and $\Hg$ is the sub-space of
$\H$ consisting of those functions of connections which depend only on
these $3N$ bits.  (Roughly, it is like focusing on only $3N$
components of a vector with an infinite number of components and
considering functions which depend only on these $3N$ components,
i.e., are constants along the orthogonal directions.) To get the full
information, we need all possible graphs. Thus, a function of
connections in $\H$ can be specified by fixing a function in $\Hg$ for
{\it every} graph $\g$ in the physical space. Of course, since two
distinct graphs can share edges, the collection of functions on $\Hg$
must satisfy certain consistency conditions. These lie at the
technical heart of various constructions and proofs.

The fact that $\H$ is the (projective) limit of $\Hg$ breaks up any
given problem in quantum geometry into a set of problems in quantum
mechanics. Thus, for example, to define operators on $\H$, it suffices
to define a {\it consistent family of} operators on $\Hg$ for each
$\g$. This makes the task of defining geometric operators feasible. I
want to emphasize, however, that the introduction of graphs is only
for technical convenience. Unlike in lattice gauge theory, we are not
{\it defining} the theory via a continuum limit (in which the lattice
spacing goes to zero.) Rather, the full Hilbert space $\H$ of the
continuum theory is already well-defined. Graphs are introduced only
for practical calculations. Nonetheless, they bring out the
one-dimensional character of quantum states/excitations of geometry:
It is because `most' states in $\H$ can be realized as elements of
$\Hg$ for some $\g$ that quantum geometry has a `polymer-like'
character.

Let me now outline the result of applying this procedure for geometric
operators. Suppose we are given a surface $S$, defined in local
coordinates by $x_3 = {\rm const}$. The classical formula for the area
of the surface is: $A_S = \int d^2x \sqrt{E^3_i E^3_i}$, where $E^3_i$
are the third components of the vectors $\vec{E}_i$. As is obvious,
this expression is non-polynomial in the basic variables
$\vec{E}_i$. Hence, off-hand, it would seem very difficult to write
down the corresponding quantum operator. However, thanks to the
background independent functional calculus, the operator can in fact
be constructed rigorously. 

To specify its action, let us consider a state which belongs to $\Hg$
for {\it some} $\g$. Then, the action of the final, regularized
operator $\hat{A}_S$ is as follows. If the graph has no intersection
with the surface, the operator simply annihilates the state. If there
are intersections, it acts at each intersection via the familiar
angular momentum operators associated with $SU(2)$. {\it This simple
form is a direct consequence of the fact that we do not have a
background geometry}: given a graph and a surface, the diffeomorphism
invariant information one can extract lies in their intersections.  To
specify the action of the operator in detail, let me suppose that the
graph $\g$ has $N$ edges. Then the state $\Psi$ has the form:
$\Psi(\vec{A}_i) = \psi(g_1, ...g_N)$ for some function $\psi$ of the
$N$ variables $g_1, ...,g_N$, where $g_k$ ($\in SU(2)$) denotes the
spin-rotation that a chiral fermion undergoes if parallel transported
along the $k$-th edge using the connection $\vec{A}_i$.  Since $g_k$
represent the possible rotations of spins, angular momentum operators
have a natural action on them. In terms of these, we can introduce
`vertex operators' associated with each intersection point $v$ between
$S$ and $\g$:
\be
\hat{O}_v\,\cdot\, \Psi (A) = \sum_{I, L} k(I, L) \vec{J}_I\cdot \vec{J}_L
\,\cdot\, \psi(g_1, ..., g_N)
\ee
where $I, L$ run over the edges of $\gamma$ at the vertex $v$, $k(I,J)
= 0, \pm 1$ depending on the orientation of edges $I, L$ at $v$, and
$\vec{J}_I$ are the three angular momentum operators associated with
the $I$-th edge. (Thus, $\vec{J}_I$ act only on the argument $g_I$ of
$\psi$ and the action is via the three left invariant vector fields on
$SU(2)$.) Note that the {\it the vertex operators resemble the
Hamiltonian of a spin system, $k(I,L)$ playing the role of the
coupling constant.}  The area operator is just a sum of the
square-roots of the vertex operators:
\be 
\hat{A}_S = \frac{G\hbar}{2c^3}\,\, \sum_v\,\, |O_v|^{\textstyle{1\over 2}}
\ee
Thus, the area operator is constructed from angular momentum-like
operators. Note that the coefficient in front of the sum is just
$\frac{1}{2}\ell_P^2$, the square of the Planck length. This fact will be 
important later.

Because of the simplicity of these operators, their complete spectrum
--i.e., full set of eigenvalues-- is known explicitly: Possible
eigenvalues $a_S$ are given by
\be
a_S = {\Pl^2\over 2}\, \sum_{v}
\Big[2\jd_v(\jd_v+1) + 2\ju_v(\ju_v+1) -
\jdu_v (\jdu_v+1)\Big]^{1\over 2} 
\ee 
where $v$ labels a finite set of points in $S$ and $\jd, \ju$ and
$\jdu$ are non-negative half-integers assigned to each $v$, subject
to the usual inequality
\be
\jd + \ju \ge \jdu \ge |\jd - \ju|\, .  
\ee 
from the theory of addition of angular momentum in elementary quantum
mechanics.  Thus the entire spectrum is discrete; {\it areas are
indeed quantized!} This discreteness holds also for the length and the
volume operators. Thus the expectation that the continuum picture may
break down at the Planck scale is borne out fully. Quantum geometry is
{\it very} different from the continuum picture. This may be the
fundamental reason for the failure of perturbative approaches to
quantum gravity.

Let us now examine a few properties of the spectrum. The lowest
eigenvalue is of course zero.  The next lowest eigenvalue may be
called the {\it area gap}. Interestingly, area-gap is sensitive to the
topology of the surface $S$. If $S$ is open, it is
$\textstyle{\sqrt{3}\over 4}\ell_P^2$. If $S$ is a closed surface
--such as a 2-torus in a 3-torus-- which fails to divide the spatial
3-manifold into an `inside' and an `outside' region, the gap turns out
to be larger, $\textstyle{2\over 4}\ell_P^2$. If $S$ is a closed
surface --such as a 2-sphere in $R^3$-- which divides space into an
`inside' and an `outside' region, the area gap turns out to be even
larger; it is $\textstyle{2\sqrt{2}\over 4}\ell_P^2$. Another
interesting feature is that in the large area limit, the eigenvalues
crowd together. This follows directly from the form of eigenvalues
given above. Indeed, one can show that for large eigenvalues $a_S$,
the difference $\Delta a_S$ between consecutive eigenvalues goes as
$\Delta a_S \le (exp -\sqrt{a_S/\ell_P^2}) \ell_P^2$. Thus, $\Delta
a_S$ goes to zero very rapidly. (The crowding is noticeable already
for low values of $a_S$. For example, if $S$ is open, there is only
one non-zero eigenvalue with $a_S < 0.5 {\ell_P}^2$, seven with $a_S <
\ell_P^2$ and 98 with $a_S < 2\ell_P^2$.) Intuitively, this explains
why the continuum limit works so well.

\subsection{Physical consequences: details matter!}

However, one might wonder if such detailed properties of geometric
operators can have any `real' effect. After all, since the Planck
length is so small, one would think that the classical and
semi-classical limits should work irrespective of, e.g., whether or
not the eigenvalues crowd. For example, let us consider not the most
general eigenstates of the area operator $\hat{A}_S$ but --as was
first done in the development of the subject-- the simplest ones.
These correspond to graphs which have simplest intersections with $S$.
For example, $n$ edges of the graph may just pierce $S$, each one
separately, so that at each one of the $n$ vertices there is just a
straight line passing through. For these states, the eigenvalues are
$a_S = (\sqrt{3}/2) n\ell_P^2$. Thus, here, the level spacing $\Delta
a_S$ is uniform, like that of the Hamiltonian of a simple harmonic
oscillator.  If we restrict ourselves to these simplest eigenstates,
even for large eigenvalues, the level spacing does not go to zero.
Suppose for a moment that this is the {\it full} spectrum of the area
operator.  wouldn't the semi-classical approximation still work since,
although uniform, the level-spacing is so small?

Surprisingly, the answer is in the negative! What is perhaps even more
surprising is that the evidence comes from unexpected quarters: the
Hawking evaporation of {\it large} black holes. More precisely, we
will see that if $\Delta a_S$ had failed to vanish sufficiently fast,
the semi-classical approximation to quantum gravity, used in the
derivation of the Hawking process, must fail in an important way.
The effects coming from area quantization would have implied that even
for large macroscopic black holes of, say, a thousand solar masses, we
can not trust semi-classical arguments. 

Let me explain this point in some detail. The original derivation of
Hawking's was carried out in the framework of quantum field theory in
curved space-times which assumes that there is a specific underlying
continuum space-time and explores the effects of curvature of this
space-time on quantum matter fields. In this approximation, Hawking
found that the classical black hole geometries are such that there is
a spontaneous emission which has a Planckian spectrum at infinity.
Thus, black-holes, seen from far away, resemble black bodies and the
associated temperature turns out to be inversely related to the mass
of the hole. Now, physically one expects that, as it evaporates, the
black hole must lose mass. Since the radius of the horizon is
proportional to the the mass, the area of the horizon must decrease.
Thus, to describe the evaporation process adequately, we must go
beyond the external field approximation and take in to account the
fact that the underlying space-time geometry is in fact
dynamical. Now, if one treated this geometry classically, one would
conclude that the process is continuous. However, since we found that
the area is in fact quantized, we would expect that the black hole
evaporates in discrete steps by making a transition from one area
eigenvalue to another, smaller one. The process would be very similar
to the way an excited atom descends to its ground state through a
series of discrete transitions.

Let us look at this process in some detail. For simplicity let us use
units with $c\! =\! 1$.  {\it Suppose, to begin with, that the level
spacing of eigenvalues of the area operator is the naive one,
i.e. with} $\Delta a_S = (\sqrt{3}/2) \ell_P^2$.  Then, the
fundamental theory would have predicted that the smallest frequency,
$\omega_o$, of emitted particles would be given by $\hbar \omega_o$ and
the smallest possible change $\Delta M$ in the mass
of the black hole would be given by $\Delta M = \hbar \omega_o$.
Now, since the area of the horizon goes as $A_H \sim
G^2M^2$, we have $\Delta M \sim \Delta a_H/2G^2 M \sim \ell_P^2/G^2M$.
Hence, $\hbar \omega_o \sim \hbar/GM$.  Thus, the `true' spectrum
would have emission lines only at frequencies $\omega = N \omega_o$,
for $N =1,2, ...$ corresponding to transitions of the black hole
through $N$ area levels. How does this compare with the Hawking
prediction? As I mentioned above, according to Hawking's
semi-classical analysis, the spectrum would be the same as that of a
black-body at temperature $T$ given by $kT \sim \hbar/GM$, where $k$
is the Boltzmann constant.  Hence, the peak of this spectrum would
appear at $\omega_{p}$ given by $\hbar \omega_p \sim kT \sim
\hbar/GM$. But this is precisely the order of magnitude of the minimum
frequency $\omega_o$ that would be allowed if the area spectrum were
the naive one. Thus, in this case, a more fundamental theory would
have predicted that the spectrum would {\rm not} resemble a black body
spectrum.  The most probable transition would be for $N=1$ and so the
spectrum would be peaked at $\omega_p$ as in the case of a black
body. However, there would be no emission lines at frequencies low
compared with $\omega_p$; this part of the black body spectrum would
be simply absent. The part of the spectrum for $\omega > \omega_p$
would also not be faithfully reproduced since the discrete lines with
frequencies $N\omega_o$, with $N=1,2,...$ would {\it not} be
sufficiently near each other --i.e. crowded-- to yield an
approximation to the continuous black-body spectrum.

The situation is completely different for the correct, full spectrum
of the area operator if the black hole is macroscopic, i.e., large.
Then, as I noted earlier, the area eigenvalues crowd and the level
spacing goes as $\Delta a_H \le (\exp -\sqrt{a_H/\ell_P^2}) \ell_P^2$.
As a consequence, as the black hole makes transition from one area
eigenvalue to another, it would emit particles at frequencies equal to
or larger than $\sim \omega_p \exp -\sqrt{a_H/\ell_P^2}$. Since for a
macroscopic black-hole the exponent is very large (for a solar mass
black-hole it is $\sim 10^{38}$!) the spectrum would be
well-approximated by a continuous spectrum and would extend well below
the peak frequency. Thus, the precise form of the area spectrum
ensures that, for large black-holes, the potential problem with
Hawking's semi-classical picture disappears. Note however that as the
black hole evaporates, its area decreases, it gets hotter and
evaporates faster. Therefore, a stage comes when the area is of the
order of $\ell_P^2$. Then, there {\it would} be deviations from the
black body spectrum. But this is to be expected since in this extreme
regime one does not expect the semi-classical picture to continue to
be meaningful.

This argument brings out an interesting fact. There are several
iconoclastic approaches to quantum geometry in which one simply begins
by postulating that geometric quantities should be quantized. Then,
having no recourse to first principles from where to derive the
eigenvalues of these operators, one simply postulates them to be
multiples of appropriate powers of the Planck length. For area then,
one would say that the eigenvalues are integral multiples of
$\ell_P^2$. The above argument shows how this innocent looking
assumption can contradict semi-classical results {\it even for large
black holes}. In the present approach, we did not begin by postulating
the nature of quantum geometry. Rather, we {\it derived} the spectrum
of the area operator from first principles. As we see, the form of
these eigenvalues is rather complicated and could not have been
guessed a priori.  More importantly, the detailed form does carry rich
information and in particular removes the conflict with semi-classical
results in macroscopic situations.

\subsection{Current and future directions}

Exploration of quantum Riemannian geometry continues. Last year, it
was found that geometric operators exhibit certain unexpected
non-commutativity. This reminds one of the features explored by Alain
Connes in his non-commutative geometry. Indeed, there are several
points of contact between these two approaches. For instance, the
Dirac operator that features prominently in Connes' theory is closely
related to the connection $\vec{A}_i$ used here. However, at a
fundamental level, the two approaches are rather different. In Connes'
approach, one constructs a non-commutative analog of entire
differential geometry. Here, by contrast, one focuses only on
Riemannian geometry; the underlying manifold structure remains
classical. In three space-time dimensions, it is possible to get rid
of this feature in the final picture and express the theory in purely
combinatorial fashion. Whether the same will be possible in four
dimensions remains unclear. However, combinatorial methods continue to
dominate the theory and it is quite possible that one would again be
able to present the final picture without any reference to an
underlying smooth manifold.

Perhaps the most striking application of quantum geometry has been to
black hole thermodynamics. We saw in section 3.3 that the Hawking
process provides a non-trivial check on the level spacing of the
eigenvalues of area operators. Conversely, the discrete nature of
these eigenvalues provides a statistical mechanical explanation of
black hole entropy. To see this, first recall that for familiar
physical systems ---such as a gas, a magnet, or a black body--- one
can arrive at the expression of entropy by counting the number of
micro-states. The counting in turn requires one to identify the
building blocks that make up the system. For a gas, these are atoms;
for a magnet, electron spins and for the radiation field in a black
body, photons. What are the analogous building blocks for a large
black hole?  They {\it can not} be gravitons because the gravitational
fields under consideration are static rather than
radiative. Therefore, the elementary constituents must be
non-perturbative in nature. In our approach they turn out to be
precisely the quantum excitations of the geometry of the black hole
horizon. The polymer-like one dimensional excitations of geometry in
the bulk pierce the horizon and endow it with its area. It turns out
that, for a given area, there are a specific number of permissible
bulk states and for each such bulk state, there is a precise number of
permissible surface states of the intrinsic quantum geometry of the
horizon. Heuristically, the horizon resembles a pinned balloon
---pinned by the polymer geometry in the bulk--- and the surface
states describe the permissible oscillations of the horizon subject to
the given pinning. A count of all these quantum states provides, in
the usual way, the expression of the black hole entropy.
 
Another promising direction for further work is construction of better
candidates for `weave states', the non-linear analogs of coherent
states approximating smooth, macroscopic geometries. Once one has an
`optimum' candidate to represent Minkowski space, one would develop
quantum field theory on these weave quantum geometries. Because the
underlying basic excitations are one-dimensional, the `effective
dimension of space' for these field theories would be less than
three. Now, in the standard continuum approach, we know that quantum
field theories in low dimensions tend to be better behaved because
their ultra-violet problems are softer. Hence, there is hope that
these theories will be free of infinities. If they are renormalizable
in the continuum, their predictions at large scales can not depend on
the details of the behavior at very small scales.  Therefore, one
might hope that quantum field theories on weaves would not only be
finite but also agree with the renormalizable theories in their
predictions at the laboratory scale.

A major effort is being devoted to the task of formulating and solving
quantum Einstein's equations using the new functional calculus. Over
the past two years, there have been some exciting developments in this
area. The methods developed there seem to be applicable also to
supergravity theories. In the coming years, therefore, there should be
much further work in this area. More generally, since quantum geometry
does not depend on a background metric, it may well have other
applications.  For example, it may provide a natural arena for other
problem such as that of obtaining a background independent formulation
of string theory.

So far, I have focussed on theoretical ideas and checks on them have
come from considerations of consistency with other theoretical ideas,
e.g., those in black hole thermodynamics.  What about experimental
tests of predictions of quantum geometry?  An astonishing recent
development suggests that direct experimental tests may become
feasible in the near future. I will conclude with a summary of the
underlying ideas. The approach one takes is rather analogous to the
one used in proton decay experiments. Processes potentially
responsible for the decay come from grand unified theories and the
corresponding energy scales are very large, $10^{15}$ GeV ---only four
orders of magnitude below Planck energy. There is no hope of achieving
these energies in particle accelerators to actually create in large
numbers the particles responsible for the decay. Therefore the decays
are very rare. The strategy adopted was to carefully watch a {\it
very} large number of protons to see if one of them decays. These
experiments were carried out and the (negative) results actually ruled
out some of the leading candidate grand unified theories. Let us
return to quantum geometry. The naive strategy of accelerating
particles to Planck energy to directly `see' the Planck scale geometry
is hopeless.  However, as in proton decay experiments, one can let
these minutest of effects accumulate till they become measurable. The
laboratory is provided by the universe itself and the signals are
generated by the so-called $\gamma$-ray bursts. These are believed to
be of cosmological origin. Therefore, by the time they arrive on
earth, they have traveled extremely large distances. Now, if the
geometry is truly quantum mechanical, as I suggested, the propagation
of these rays would be slightly different from that on a continuum
geometry. The difference would be minute but could accumulate on
cosmological distances. Following this strategy, astronomers have
already put some interesting limits on the possible `gaininess' of
geometry. Now the challenge for theorists is to construct realistic
weave states corresponding to the geometry we observe on cosmological
scales, study in detail propagation of photons on them and come up
with specific predictions for astronomers. The next decade should
indeed be very exciting!

\bigskip

{\bf Acknowledgments} The work summarized here is based on
contributions from many researchers especially John Baez, Alejandro
Corichi, Roberto DePitri, Rodolfo Gambini, Chris Isham, Junichi
Iwasaki, Jerzy Lewandowski, Renate Loll, Don Marolf, Jose Mourao,
Jorge Pullin, Thomas Thiemann, Carlo Rovelli, Steven Sawin, Lee Smolin
and Jos\'e-Antonio Zapata.  Special thanks are due to Jerzy
Lewandowski for long range collaboration.  This work was supported in
part by the NSF Grant PHY95-14240 and by the Eberly fund of the
Pennsylvania State University.

\bigskip\bigskip


\end{document}